\documentclass[twocolumn]{aastex61}
\usepackage{natbib}
\usepackage{epsfig,graphicx,times}
\usepackage{subfigure}
 \usepackage{CJK}
\begin{document}

\title{Giant planets around FGK stars form probably through core accretion
}
\author{Wei Wang}
\affiliation{Key Laboratory of Optical Astronomy, National Astronomical Observatories,
Chinese Academy of Sciences, Beijing 100101, China}
\affiliation{ Chinese Academy of Sciences South America Center for Astronomy, National Astronomical Observatories, Chinese Academy of Sciences, Beijing 100101, China}
\author{Liang Wang}
\affiliation{Max-Planck-Institut f\"ur Extraterrestrische Physik, Giessenbachstrasse, D-85748 Garching, Germany}
\affiliation{Key Laboratory of Optical Astronomy, National Astronomical Observatories,
Chinese Academy of Sciences, Beijing 100101, China}
\author{Xiang Li}
\affiliation{Key Laboratory of Optical Astronomy, National Astronomical Observatories,
Chinese Academy of Sciences, Beijing 100101, China}
\author{Yuqin Chen}
\affiliation{Key Laboratory of Optical Astronomy, National Astronomical Observatories,
Chinese Academy of Sciences, Beijing 100101, China}
\affiliation{School of Astronomy and Space Science, University of Chinese
Academy of Sciences, Beijing 100049, China}
\author{Gang Zhao}
\affiliation{Key Laboratory of Optical Astronomy, National Astronomical Observatories,
Chinese Academy of Sciences, Beijing 100101, China}
\affiliation{School of Astronomy and Space Science, University of Chinese
Academy of Sciences, Beijing 100049, China}

\correspondingauthor{Wei Wang, Gang Zhao}
\email{wangw@nao.cas.cn, gzhao@nao.cas.cn}

 \shorttitle{Giant planets probably form through core-accretion} 
 \shortauthors{Wei Wang et al. }


\begin{abstract}

We present a statistical study of the planet-metallicity (P-M) correlation, by
comparing the 744 stars with candidate planets (SWPs) in the Kepler field which
have been observed with LAMOST, and a sample of distance-independent, 
fake ``twin'' stars in the Kepler field with no planet reported (CKSNPs) yet.
With the well-defined and carefully-selected large samples, we find for the
first time a turn-off P-M correlation of $\Delta$[Fe/H]$_{\rm SWPs-SNPs}$,
which in average increases from $\sim0.00\pm0.03$\,dex to $0.06\pm0.03$\,dex,
and to $0.12\pm0.03$ for stars with Earth, Neptune, Jupiter-sized planets
successively, and then declines to $\sim-0.01\pm0.03$\,dex for more massive
planets or brown dwarfs. Moreover, the percentage of those systems with
positive $\Delta$[Fe/H] has the same turn-off pattern. We also find 
FG-type stars follow this general trend, but K-type stars are different.
Moderate metal enhancement ($\sim0.1-0.2$\,dex) for K-type stars with planets of
radii between 2 to 4 $R_{\oplus}$ as compared to CKSNPs is observed, which
indicates much higher metallicities are required for Super-Earths,
Neptune-sized planets to form around K-type stars. We point out that the P-M
correlation is actually metallicity-dependent, i.e., the correlation is
positive at solar and super-solar metallicities, and negative at subsolar
metallicities. No steady increase of $\Delta$[Fe/H] against planet sizes is
observed for rocky planets, excluding the pollution scenario as a major
mechanism for the P-M correlation. All these clues suggest that giant planets
probably form differently from rocky planets or more massive planets/brown
dwarfs, and the core-accretion scenario is highly favoured, and high
metallicity is a prerequisite for massive planets to form. 

\end{abstract}

\section{Introduction}

The dependences of planet occurrence rate on host star properties provide
important input for our understanding of planet formation and evolution,
especially the well-known planet-metallicity (P-M) correlation as discovered by
\cite{Gonzalez1997} for giant planets, i.e., the planet occurrence is higher
around metal-rich stars. This correlation may be the signature of
self-pollution during the planet formation process \citep[e.g.][]{Gonzalez1997,
Murray2002a, Murray2002b}, following the planet migration description of
\cite{Lin1996}, or indicates that high metallicity is a prerequisite for the
formation of gas giants \citep[e.g.]{FV2005}, as demanded by the core accretion
scenario~\cite[e.g.][]{IL2005}. We note that selection effects might be a third
factor for the correlation at least for the studies concentrated on planets
found via high-resolution spectroscopic radial velocity methods, where the
sample of stars with planets is naturally biased to high metallicity and nearby
bright stars. However, this effect is less significant for the samples
with planets discovered through other techniques, e.g. the transit method.

The NASA {\textit Kepler} mission \citep{Borucki2011} has had great success in
finding transiting exoplanet candidates. More than 3500 planet candidates have
been discovered during its 4.5-year mission \citep{Batalha2013, Burke2013}.
This mission has a unique power to make it possible to study the planet
occurrence rate and especially $\eta_{\rm Earth}$, the fraction of Sun-like
stars harboring Earth-like planets, in a reliable way. More importantly, the
mission provides a high confidence for either the detection of planets (false
positives $\sim7\%-18\%$ depending on the radii of planets, \citealt{MJ2011,
Fressin2013, Morton2016}) or the number of stars being searched but without
known planets; The latter is guaranteed by the high precision and time coverage
of {\textit Kepler} photometry, which has never been achieved in the past.
Afterwards, planet occurrence rate has been extensively studied \citep{CS2011,
Howard2012, Fressin2013, DC2013}, and its dependency on metallicity (P-M
correlation) has also been tackled, for example, by \cite{Buchhave2012},
\cite{Everett2013}, \cite{WF2013}, \cite{Buchhave2014} and
\cite{Schlaufman2015}. 

\begin{figure*}%
 \fbox{\includegraphics[width=.30\textwidth]{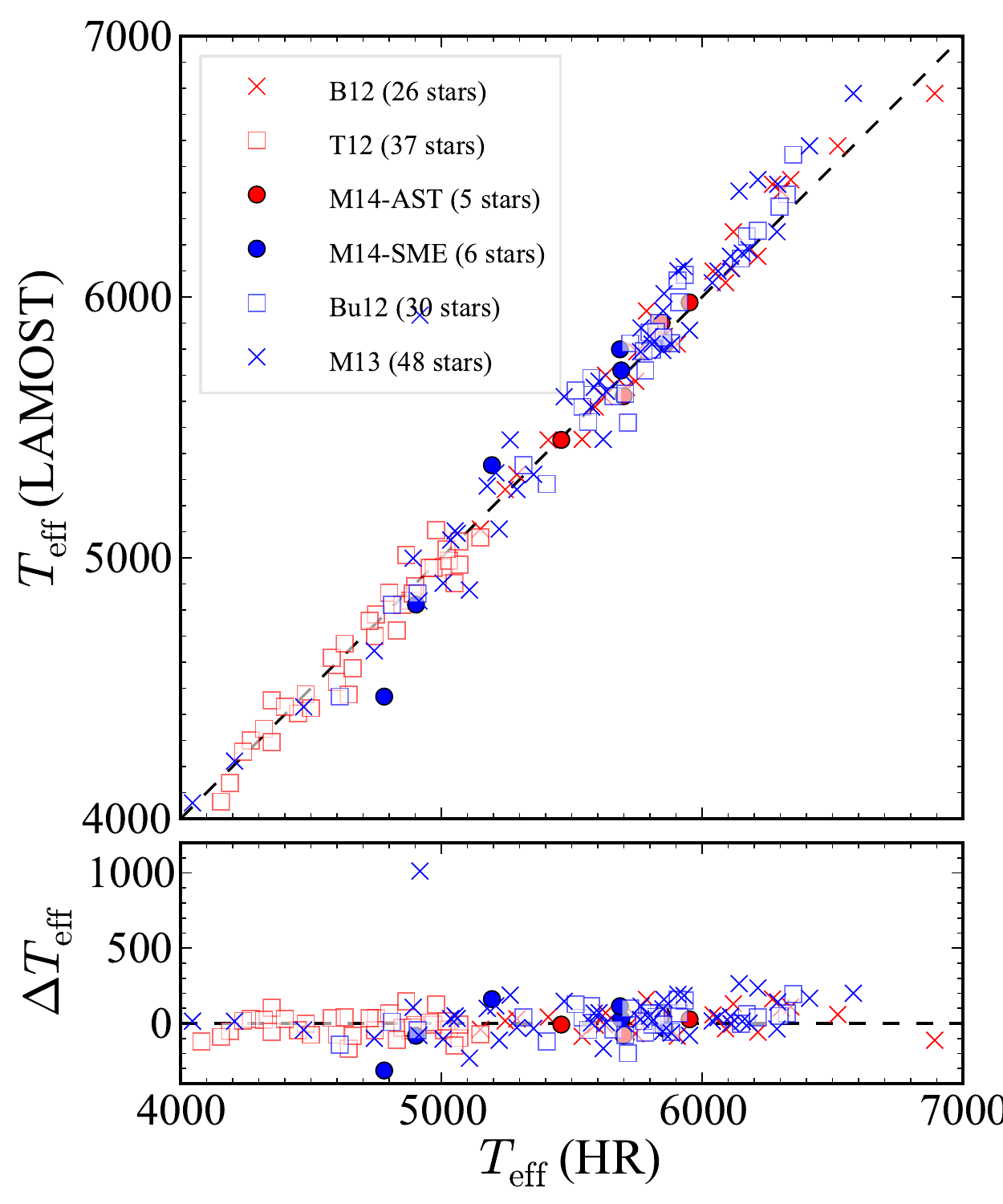} }%
\qquad
 \fbox{\includegraphics[width=.30\textwidth]{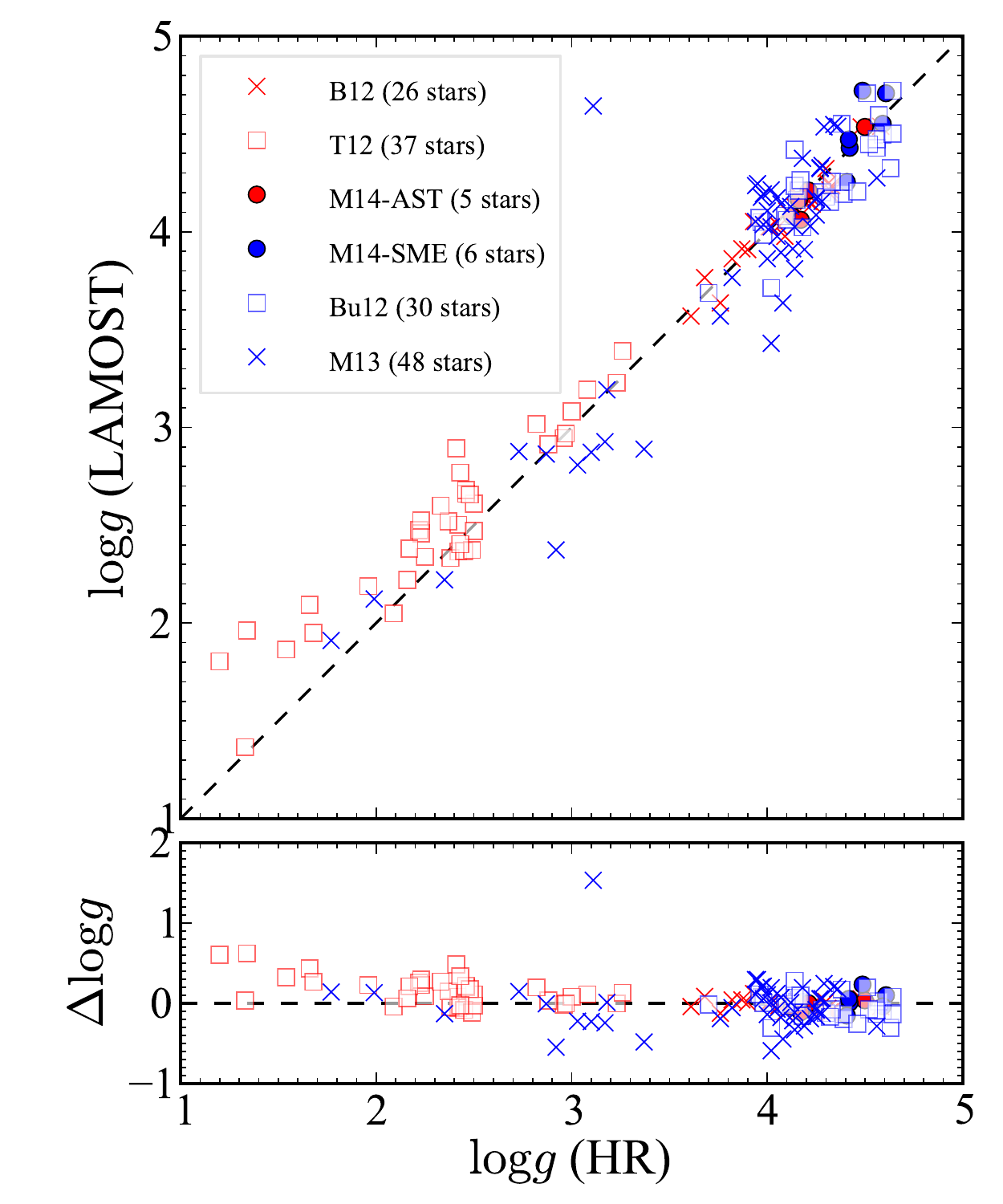} }%
\qquad
 \fbox{\includegraphics[width=.30\textwidth]{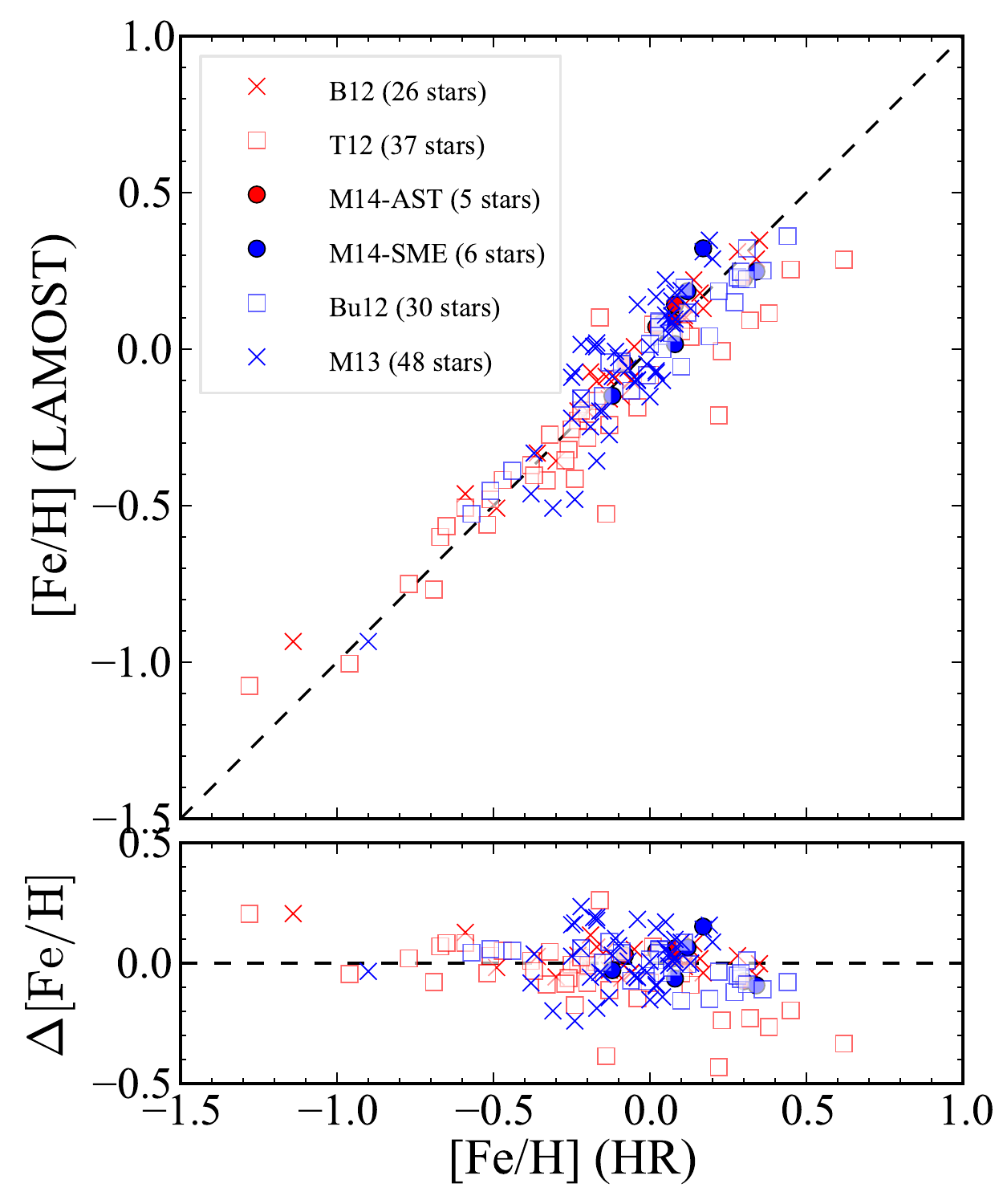} }%
\caption{\label{fig:AP_comparisons} Comparisons of $T_{\rm eff}$ ({\bf Left})
and [Fe/H] ({\bf Right}) between those from LAMOST DR3 and those from recent
high-resolution spectroscopic studies by \citealt{Bruntt2012},
\citealt{Thygesen2012}, \citealt{Molenda2013},
\citealt{Buchhave2012},\citealt{Marcy2014}.}%
\end{figure*}

\cite{Buchhave2012} measured precise stellar parameters and metallicity [m/H]
for a sample of 152 \textit{Kepler} planet-host stars using their stellar
parameters classification (SPC) tool (Supplementary information of
\citealt{Buchhave2012}). They used a set of high-resolution low-to-moderate
($\sim13-312$) SNR echelle spectra, and fit them in the wavelength ranged
between 5050 and 5360\,{\AA} to a library grid of synthetic model spectra.
They found a positive correlation between the planet occurrence rate and
stellar metallicity for giant planets, i.e., the planet-metallicity (P-M)
correlation. However, the metallicity pattern for stars hosting planets with
radii smaller than 4$R_{\oplus}$ does not clearly reveal any correlations.
Instead, a wide range of metallicity distribution was observed in their sample.
\cite{Everett2013} derived stellar parameters and [Fe/H] for 220 faint {\it
Kepler} planet candidate host stars, by fitting model spectra published by
\cite{Coelho2005} to moderate-resolution (R$\sim3000$) optical spectra covering
$3800-4900$\AA. They concluded that the frequency of large planets in the
sample depends on host star metallicity, similar to that found for the sample
of brighter KOI stars by \cite{Buchhave2012}. Later on, by employing stellar
and orbital parameters from the KOI
website$\footnote{$^1$http://exoplanetarchive.ipac.caltech.edu}$, and stellar
metallicities from \textit{Kepler} Input Catalog (KIC, \citealt{Brown2011}),
\cite{WF2013} investigated this question in a better sample including only
multi-planet systems which are claimed to have a substantially lower false
positive rate \citep{Lissauer2012}. They confirmed the positive P-M correlation
for gas giant planets with radius between 5$R_{\oplus}$ and 22$R_{\oplus}$, and
it holds for Neptune-size planets with radius between 2$R_{\oplus}$ and
5$R_{\oplus}$, however not for Earth-sized planets with radius less than
2$R_{\oplus}$. \cite{Buchhave2014} used the same method to derive stellar
parameters for 405 stars orbited by 600 exoplanet candidates, and they claimed
that the planets can be categorised into three regimes defined by
statistically distinct metallicity regions, reflecting the different formation
scenarios of rocky planets, gas dwarf planets and gas/icy giant planets.

More recently, \cite{WF2015} and \cite{BL2015} used a same sample of 405 stars
with transiting planets but different reference stellar samples to explore the
P-M correlation. While \cite{WF2015} used a sample of Solar-like Kepler stars
with no known planets as the reference sample, \cite{BL2015} employed the 518 dwarf
stars from the asteroseismic sample~\citep{Chaplin2014} for comparisons.
Interestingly and instructively, the former exercise detected the P-M
correlation for terrestrial planets, while the later reported a null detection
for them. This indicates it is crucial to choose and refine an appropriate and
unbiased reference sample, and to have reliable or at least consistent
determinations of stellar metallicities.

We note that in \cite{WF2013}, the stellar metallicity [Fe/H] is taken from
\textit{Kepler} Input Catalog (KIC, \citealt{Brown2011}), which is known to be
of large uncertainties of $\sim0.4$\,dex, and is recently found to
underestimate both the true metallicity and dynamic range \citep{Everett2013,
Dong2014}, thanks to Data Release 1 (DR1, \citealt{Luo2012}) of the Large Sky
Area Multi-Object Fiber Spectroscopic Telescope (LAMOST, a. k. a., Guoshoujing
telescope, \citealt{Cui2012, Zhaogang2012}). In 2016, LAMOST DR3 (hereafter
LMDR3) has been released, includes 3,177,995 FGK-type stars with stellar
atmospheric parameters (APs) automatically determined consistently using the
LAMOST Stellar Parameter Pipeline (LSPP, \citealt{Wu2011}), at precisions of
110\,K, 0.19\,dex and 0.11\,dex for stellar effective temperature Teff, gravity
log g and metallicity [Fe/H], respectively\citep{Wu2014}. We further examined
LSPP $T_{\rm eff}$ and [Fe/H] with those determined from high resolution
spectroscopic data in three recent work
\citep{Bruntt2012,Thygesen2012,Molenda2013}, and found a good agreement with a
median discrepancy of $-36\pm108$ K in $T_{\rm eff}$ and $-0.02\pm0.11$\,dex in
[Fe/H], as seen in Fig.~\ref{fig:AP_comparisons}. In \citealt{WangL2016}, we
made an extensive study of the log\,\textit{g} determinations from LSPP and
those from asteroseismology with the {\textit Kepler} data, and performed a
calibration of the former with the latter. We surprisingly found that LSPP
gives quite robust internal log\,\textit{g} in the sense that by applying a
piecewise linear functions of the difference between LSPP log\,\textit{g} and
asteroseismic log\,\textit{g} against LSPP $T_{\rm eff}$, the residual has a
close-to-zero mean value of $-0.02$ and a standard deviation of merely
0.13\,dex, with only 3\% outliers. In summary, the error budget for LSPP APs
are 108\,K, 0.13\,dex and 0.11\,dex for $T_{\rm eff}$, log\,\textit{g} and
[Fe/H], respectively. LMDR3 includes about 68,764 \textit{Kepler} stars as
part of the ``LAMOST-\textit{Kepler} project'', which aims to combine these two
unique surveys together for various studies including planetary science and
stellar physics, i.e., \textit{Kepler} provides high quality photometry and
light curves for the studies of stellar asteroseismology and transit planets,
while LAMOST can provide relatively reliable and self-consistent atmospheric
parameters (APs) for the stars rapidly. 

With the emergence of the large data set of medium-resolution spectra
produced by LAMOST, several investigations come up very recently with quite
interesting results. Employing the LAMOST-\textit{Kepler} data set,
\citet{Mulders2016} studied the dependences of the P-M correlation on orbital
period and planet size, and found that planets with orbital periods less than
ten days are more likely detected around metal-rich stars, and this trend is
most significantly for rocky planets. \cite{Dong2018} discovered a population
of short-period, Neptune-sized planets, which are similar to hot Jupiters in
the aspects that they are both preferentially orbiting metal-rich stars as
revealed by LAMOST, and are both in the mean time preferentially
single-transiting planetary systems as observed by {\textit Kepler}. 

In this study, we present our results using the ``LAMOST-\textit{Kepler}
project'' to study the planet-metallicity correlation. We will show in
Section~2 how reliable the stellar parameters given in LMDR3 are. In Section 3,
we will introduce the target sample of 744 Kepler stars with detected candidate
planets (SWPs) and two control samples bKSNP and CKSNP, and will discuss them
in details in Section~4. A brief summary will be given in the last section.

\section{The target sample and the control samples}

\begin{figure*}
\includegraphics[width=18cm,clip=true,bb=0 0 623 311]{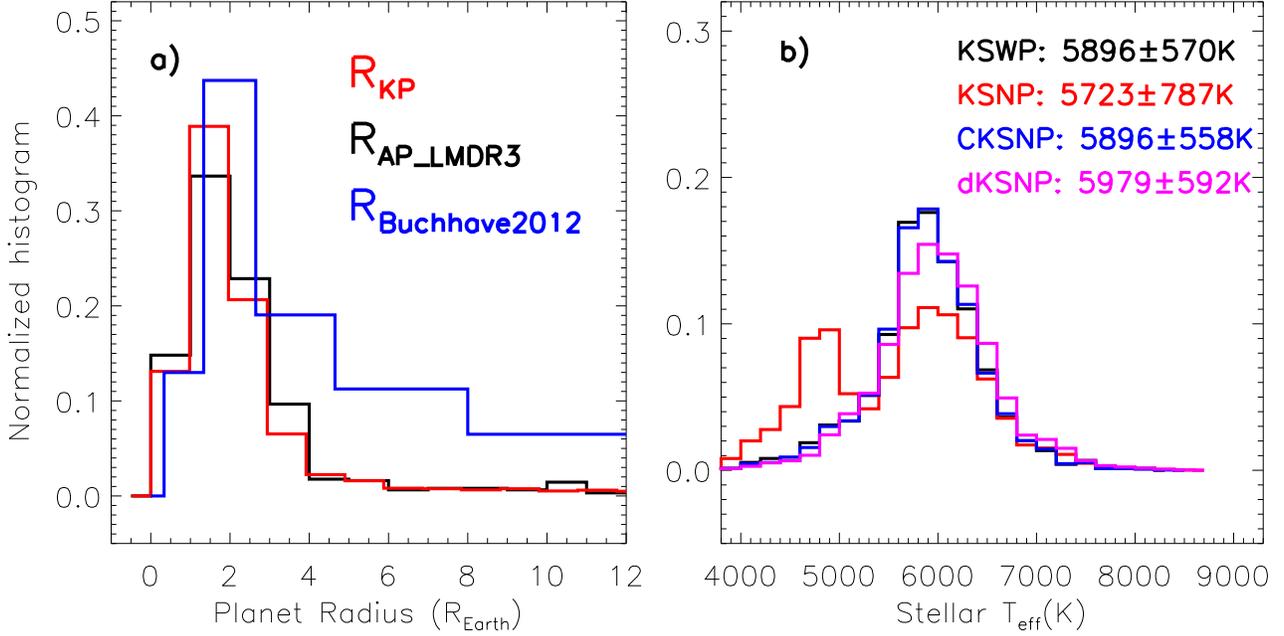}
\caption{\label{fig:rad_teff_dist} The distribution of planet radius (Left) and
stellar effective temperature $T_{\rm eff}$ (Right). {\textit Left}: red
histogram for all \textit{Kepler} candidate planets, black for all planets from
EDE found in LAMOST DR3, blue for the planets in \citet{Buchhave2012};
{\textit Right}: median values and standard deviations of $T_{\rm eff}$ for
every samples are marked. The purple histogram is the distribution of $T_{\rm
eff}$ of dwarf KSNPs with log\,$g>3.5$.
}
\end{figure*}

We cross checked LMDR3 with {\textit Kepler} candidate catalogue and Exoplanet
Data Explorer (http://exoplanets.org, hereafter EDE, \citealt{Han2014}), and
found 801 stars harbouring in total 1045 planets. They are targeted as field
stars in LAMOST Experiment for Galactic Understanding and Exploration (LEGUE)
survey \citep{Deng2012}, and the LAMOST-{\it Kepler} project \citep{DeCat2015}.
We restrain the sample into dwarf stars with log\,g$>3.5$, to avoid any
possible metal contaminations during the late stellar evolutionary stages in
(sub)giant stars. We further remove the KOI\,607 system, because recently the
companion is found to be a low-mass star instead of a planet or brown dwarf,
with $M=0.106\pm0.006\,M\odot$ measured through 2 high-precision velocimetry
measurements (\citealt{Santerne2012}). Among the rest, there are 744 {\textit
Kepler} stars with SDSS $g$ and $r$ magnitudes reported in literatures, this
assembles the {\bf SWP} sample. The planets around stars in this sample are
mostly candidate Super-Earths or Neptunes. It has a size distribution very
similar to that of the entire {\textit Kepler} candidate planet sample, as
shown in Fig.~\ref{fig:rad_teff_dist}. In contrast, the size distribution of
candidate planets in \cite{Buchhave2012} is obviously different from the entire
{\textit Kepler} planet sample (the blue histogram compares the red in
Fig.~\ref{fig:rad_teff_dist}); their sample includes slightly smaller fraction
of planets with radii below 1.3$R_{\oplus}$ ($\sim$25\%), and much higher
fraction (a factor of $\sim$2) of planets with radii larger than 4$R_{\oplus}$. 

There are {\bf 67,787} Kepler stars with no known planets that have spectra
taken by LAMOST and released in DR3. These stars form a control SNP sample,
named as {\bf KSNP}. However, we stress that for such investigations, KSNP is
not a well-defined control sample. In Fig.~\ref{fig:rad_teff_dist}b,
distributions of $T_{\rm eff}$ given by LSPP are plotted for the target SWP
sample and the three control samples. There is a clear deviation of the
distributions of SWP (black) and KSNP (red) at lower stellar $T_{\rm eff}$,
especially at $T_{\rm eff}<5300$\,K. Further investigations show that about
78\% of KSNPs within this $T_{\rm eff}$ regime are giants or sub-giants with
log\,$g<3.5$, and only $\sim$22\% of SWPs are not dwarfs. If we remove all
KSNPs with log\,$g<3.5$ (the purple histogram), the large difference at
$\sim5000$\,K disappears. It is not unexpected given that all the planets under
current study were found using the transit method, with which the detection
rates are highly dependent on the stellar radius, and therefore largely different
for giant and dwarf stars. Additionally, giants and dwarfs have statistically
different metallicity distributions. We note that thus cautions should be made
for a clean selection of the control sample for the study of the P-M
correlation, but has never done well enough in any previous studies.

The best way to tackle the P-M correlation is to compare a sample of twin
stars, with one of the twin hosting planets while the other not. In this case,
the presence of planets is the only cause of metal enrichment, or the only
result of high metallicity. To approach this, we assemble a control sample
consisting of ``composite twin'' stars, to minimize possible influences induced
by $T_{\rm eff}$, log\,\textit{g} and stellar distance (D), and to study a
``clean'' P-M relationship. For each of the 744 planet-host stars, we chose
from the KSNP sample 9 reference stars, which are the most similar to the
target star, but with no known planets. In practice, the difference between SWP
and their reference stars is represented using the distance in the {($T_{\rm
eff}$, log\,\textit{g}, dereddened $r$ magnitude) 3-D parameter space for each
SWP and its counterparts. The units for these three quantities are 100\,K,
1\,dex, 1\,mag for the parameter space distance calculations. The stars with
the smallest distances and with $r$-band signal-to-noise ratio (SNR) higher
than three were chosen. Such kind of selection process naturally results in a
very similar distribution of $T_{\rm eff}$ (Fig~\ref{fig:rad_teff_dist}b.),
log\,\textit{g} and the dereddened $r$-band magnitude of SWPs and SNPs. These
similarities will consequently lead to similarities in stellar mass ($M$),
radius ($R$) and stellar distance to first-order approximation if metallicities
do not differ significantly, as well as detection rates of planets. The total
6696 analog stars selected from LMDR3 Kepler field stars assemble the ``best''
Kepler SNP (bKSNP) sample.

For each SWP, a comparison fake star is ``created'' in AP space, with APs using
the median values of the nine most similar KSNPs. The ``composite'' fake star and
its corresponding SWP are therefore star analogs or star twins, according to
the definitions of solar twins and solar analogs~\citep{SK1998}. These 744 fake
analog stars assemble the {\bf CKSNP} sample, which will be used in the
following analysis. We further check the differences of SWPs and CKSNPs in
2MASS color spaces, namely $J-H$, $H-K{\rm s}$ and $J-K{\rm s}$. The
differences in the three colors are mostly smaller than 0.1\,mag, with standard
deviations 0.04, 0.03 and 0.05\,mag, respectively. The small discrepancies in
2MASS colors provide independent support to the similarities of SWP and CKSNP
stars. We show in Fig.\ref{fig:AP_diff} the differences of $T_{\rm eff}/100$,
log\,\textit{g} between SWP and its corresponding twin-like composite SNP.
Most of these pairs have differences $<10$\,K in $T_{\rm eff}$, and 0.1\,dex in
log\,\textit{g}. On the other hand, obvious deviations of [Fe/H] distributions
in these two samples do exist, as shown by the red circles and more clearly in
the red histogram Fig.\ref{fig:AP_diff}. At first glance, there is no clear
evidence for metal enrichments of SWPs compared to SNPs as a bulk, given their
very small difference ($-0.00\pm0.001$\,dex) of the mean [Fe/H] values.
However, the red circles in the left panel shows clearly how the [Fe/H]
discrepancy in the two samples evolve interestingly with SEQ, which is
representing the increasing indices of [Fe/H]$_{\rm SWP}$. Further discussions
about this issue will be presented in Section~\ref{sec:dis}.

We emphasize here the stellar distance is an important parameter that should be
taken into account but previously ignored when selecting control sample.
Because if we change the parameter distance between SWP and SNP samples in the
$r$ magnitude space from $2$, $1$ to $-1$ and $-2$, i.e., the control sample
stars are further and further away from us, the mean [Fe/H]$_{\rm bKSNP}$
steadily decreases from $-0.01$, $-0.03$ to $-0.06$ and $-0.07$, accordingly
(Fig.~\ref{fig:FeH_vs_Teff}), in which it is clearly shown that this trend is
also true for each $T_{\rm eff}$ bin. We emphasized here that one must be very
careful with the selection of control samples for the study of the P-M
correlation and plane frequency. With an inappropriate control sample, an
inadequate, or wrong conclusion might be drawn. In our case, $\sim$97\% of the
bKSNP star have a $r$ mag difference from that of SWP within 0.25\,mag, and
$\sim$79\% of them within 0.10\,mag, which corresponds to a difference of 12\%
and 5\% in distance assuming a same stellar intrinsic luminosity, respectively.
Therefore, for each SWP, its corresponding 9 bKSNP reference stars are very
nearby and similar stars, and thus much suitable for such a study, when
compared with other similar investigations. 

\begin{figure*}
\centering
\includegraphics[width=18cm,clip=true, bb=0 0 680 340]{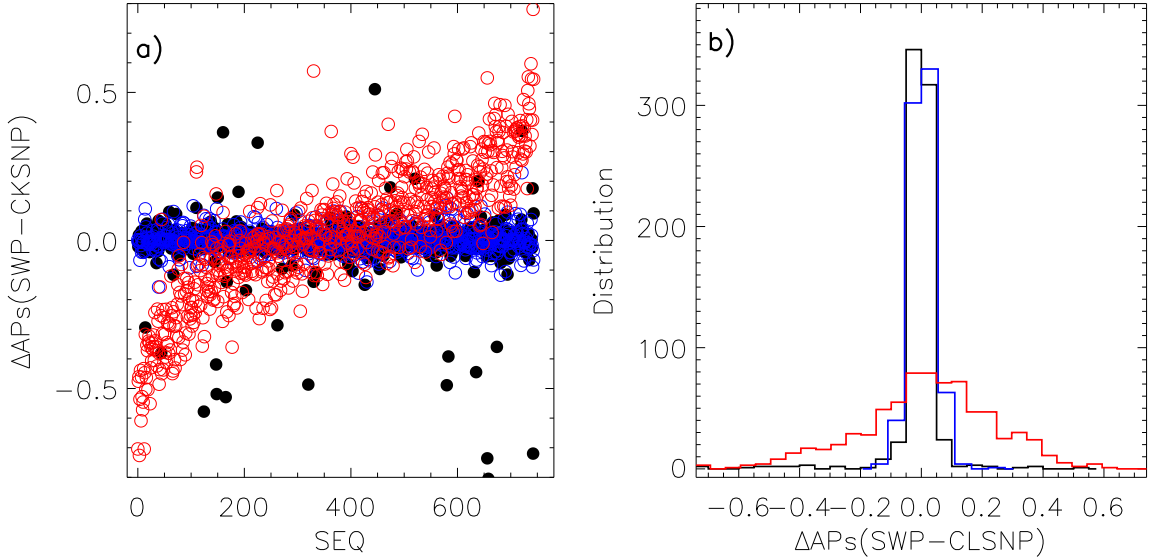}
\caption{\label{fig:AP_diff} The differences of $T_{\rm eff}/100$,
log\,\textit{g} and [Fe/H] between SWPs and CKSNPs in blue, black and red,
respectively, shown as scatter points with increasing SWP metallicity in the
left panel and histogram in the right panel. The horizontal axis in the left
panel, SEQ, is the sequence of the SWP stars in order of increasing
metallicity.The mean values of these differences are, respectively, $-0.5$\,K,
$0.003$ and $0.02$\,dex.}
\end{figure*}

\begin{figure}
\centering
\includegraphics[width=9cm,clip=true, bb=0 0 566 566]{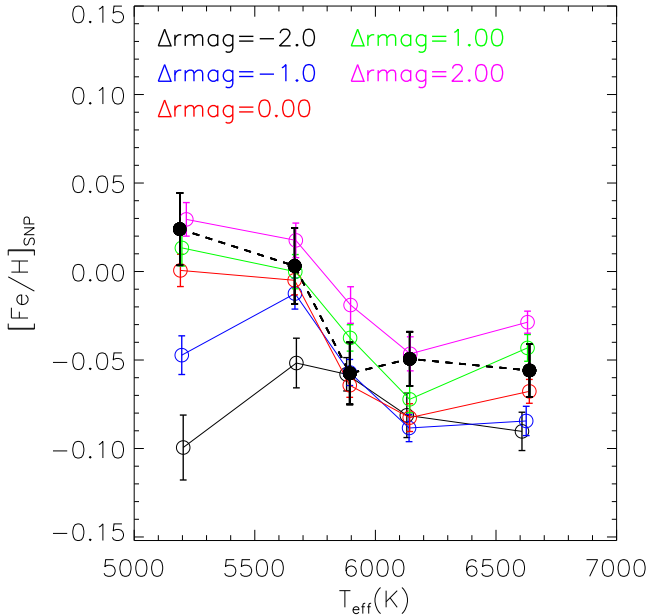}
\caption{\label{fig:FeH_vs_Teff} Average [Fe/H] in various $T_{\rm eff}$ bin as 
a function $T_{\rm eff}$. The black dashed line represents for SWPs, and the 
solid lines for CKSNPs with $\Delta\ r$ from -2.0 to 2.0\,mag, respectively. 
}
\end{figure}

The same exercise is also carried out for the entire LMDR3 SNP sample (LSNP),
consisting of about 1 million stars. We find a significant and positive large
metallicity enhancements ($\sim0.20\pm0.01$\,dex) of SWPs compared with LSNPs.
Then we realize that this might be a result of the fact that the Kepler field
stars without planets are already generally metal-rich than those in other
Galactic fields (cf. for example \citealt{Dong2014}), which means it is better
to use control stars in the Kepler field to eliminate the
metallicity-spatial-variation bias, as done in this work.

In the following, our discussion and conclusions will be mostly based on the
comparative analysis between SWP and CKSNP (the ``composite'' Kepler SNP
sample). We prefer CKSNP to dKSNP (the dwarf KSNP sample as shown in purple curve in
\ref{fig:rad_teff_dist}). The main reason is that CKSNP and SWP stars have
similar $T_{\rm eff}$ and log\,\textit{g}, and therefore the LSPP should in
principle give similar systematics and uncertainties in the determinations of
[Fe/H], and therefore comparing them will be more self-consistent and
homogeneous. Another major reason is that CKSNP and SWP have very similar
stellar distance, making them possible to have same origins and therefore
similar initial metallicities. 

\section{Discussion}
\label{sec:dis}

SWPs with giant planets have averagely higher metallicities than SNPs, as
observed, for example, by \cite{FV2005}, \cite{Neves2013}, and as interpreted
by the core accretion planet formation theory \citep[][]{IL2004,EC2010}. Such
positive relationship between planet occurrence rate and stellar metallicity
becomes partially vanish for stars hosting Neptunian-sized and even smaller
planets \citep[e.g.][]{Sousa2008, Bouchy2009, Ghezzi2010,
Sousa2011,Buchhave2012,Adibekyan2012a}. Fig.~\ref{fig:FeH_dist} shows the
[Fe/H] distributions for SWPs (the red histogram) and CKSNPs (the black one)
and bKSNPs (the blue one). The SWP, CKSNP and bKSNP samples samples have
mean [Fe/H] values of $\sim-0.03\pm0.008$, $\sim-0.05\pm0.008$ and
$\sim-0.05\pm0.003$\,dex, respectively, where the error bars are standard
errors of the means. We conclude that there is not significant difference
between the SWP and SNP samples, as shown by the red and blue histograms and
the one-component Gaussian-fitting curves in the plot.  Given the fact that
most (80\%) of our SWPs harbor planets of $R_{\rm P}<5R_{\oplus}$, the observed
simliar metallicity suggests that stars with Neptune-sized and Earth-sized
planets might not be distinguishable from those without known planets. This
discovery seems to be in line with the conclusion drawn by
\citet{Buchhave2012}} and is different from \cite{WF2015} and
\cite{Adibekyan2012a}. The latter authors carried out a uniform and detailed
abundance analysis of 12 refractory elements for a sample of 1111 FGK dwarf
stars from the HARPS GTO planet search program and found that the 26
Neptunian/super-Earth hosts have average [Fe/H] value $\sim0.23$\,dex lower
than Jovian hosts, and $\sim0.04$\,dex higher than Non-planet hosts.

Although the P-M correlation is not observed with high significance when
comparing in general our target and control samples, we will show in the
following three subsections that the so-called P-M correlation is evidently
shown. In addition, we find that the P-M correlation is dependent to
planet radius, stellar metallicity and stellar spectral types. The first
dependency, i.e., different P-M correlation at different planet radius, has
been reported and discussed extensively previously, but the last two
dependencies are both firstly reported in the current study.

\begin{figure}
\centering
\includegraphics[width=10cm,clip=true, bb=15 8 610 600]{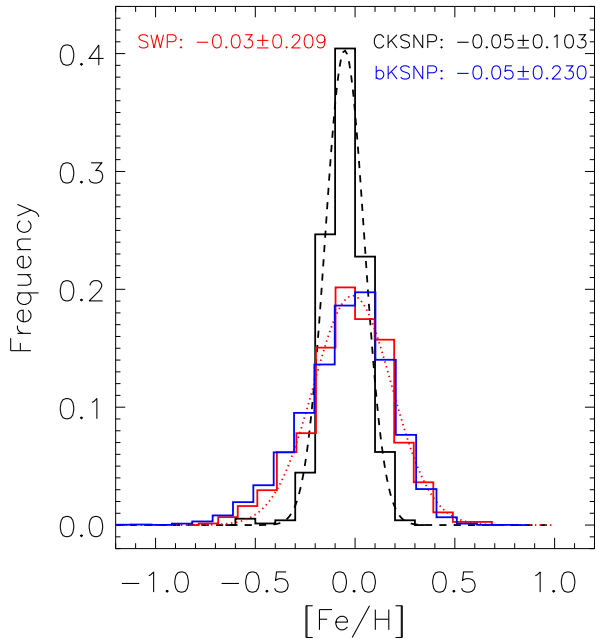}
\caption{\label{fig:FeH_dist}The histogram of [Fe/H] of CKSNPs (black), bKSNP
(blue) and SWPs (red). The black and red over-plotted dashed lines are
one-component Gaussian fits to the distributions of [Fe/H]$_{\rm CKSNP}$ and
[Fe/H]$_{\rm SWP}$, respectively. The mean metallicities and their standard
error of mean of each sample are shown in the plot.}
\end{figure}

\subsection{The P-M correlation is confirmed to depend on the planetary radius}

\begin{figure}
\centering
\includegraphics[width=9.5cm,clip=true, bb=15 8 610 550]{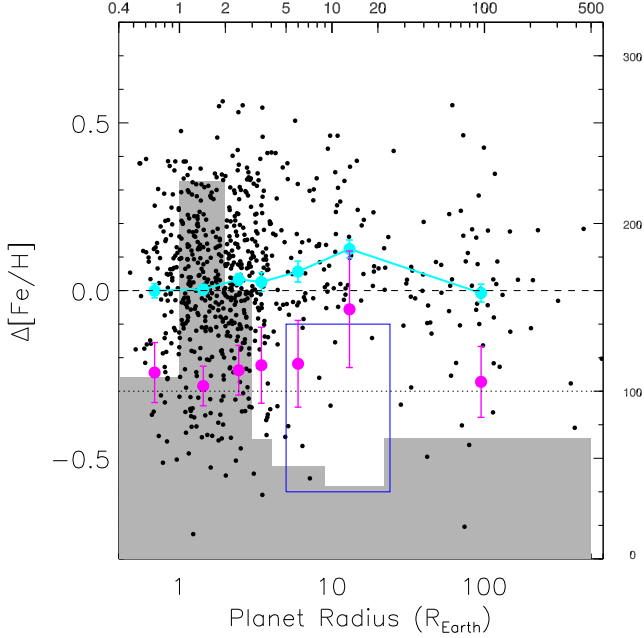}
\caption{\label{fig:dfeh2prad}$\Delta$[Fe/H] against $R_{\rm P}$ in units of
$R_{\oplus}$ for the 744 comparison pairs (black dots). The mean and standard
error of $\Delta$[Fe/H] in each of the 6 planetary radius bin are plotted in
cyan symbols with error bars. The shaded grey histogram shows the number of
planets in each bin. The blue box indicates a region that lack of stars with
$4R_{\oplus}< R_{\rm P} <24R_{\oplus}$, and $\Delta$[Fe/H] between $-0.6$ and
$-0.1$\,dex. The purple dots with error bars give the percentages and their
statistical uncertainties of metal-rich SWPs.} 
\end{figure}

In Fig.~\ref{fig:dfeh2prad}, we plot $\Delta$[Fe/H]=[Fe/H]$_{\rm
SWP}-$[Fe/H]$_{\rm CKSNP}$ against planet radius $R_{\rm P}$ in units of Earth
radius ($R_{\oplus}$) using the black filled circles for the 744 SWP$-$CKSNP
pair. Cyan points with error bars represent the average [Fe/H] in each uneven
bin, grouped in 1, 2, 3, 4, 9, 22, and 500$R_{\oplus}$, which clearly shown a
turn-off trend, peaking at $\sim9R_{\oplus}$. The bin size is indicated by the
width of the shaded grey histogram, and the histogram represents the number of
targets in each bin scaled by the right-side Y-axis. It is clearly seen that
the mean $\Delta$[Fe/H] values are close to zero for SWEPs (stars with
Earth-sized planets) with $R_{\rm P}< 2R_{\oplus}$, increase to $\sim$0.03\,dex
for stars with Super-Earths with $ 2R_{\rm P}<R_{\rm P}< 4R_{\oplus}$, and to
$0.07\pm0.03$ for SWNPs (stars with Neputian-sized planets) with
$4R_{\oplus}<R_{\rm P}<9R_{\oplus}$ and becomes evidently positive
($0.12\pm0.03$) for SWJPs (stars with Jupiter-sized planets) with
$9R_{\oplus}<R_{\rm P}<22R_{\oplus}$, and turns nearly zero ($0.02\pm0.03$) for
even larger planets or brown dwarfs. We note that the same turn-off pattern
still holds for our data set, if [Fe/H]$_{\rm SWP}$, instead of $\Delta$[Fe/H],
is to be plotted and compared. This is can be well explained by the fact that
[Fe/H]$_{\rm CKSNP}$ is more narrowly distributed compared to [Fe/H]$_{\rm
SWP}$. Furthermore, it seems that SWJPs have a high possibility to be
metal-enriched, i.e., with $\Delta$[Fe/H]$>0$. We calculate the fraction of
metal-enriched SWPs (f) in each radius bin, and the resulting percentage is
shown as the purple dots in Fig.~\ref{fig:dfeh2prad}. For reference, the
percentage of 50\% is shown as the dotted horizontal line. It is clearly shown
that SWJPs and SWNPs are both likely ($74\pm17$\% \& $58\pm12$\%) to be
metal-enriched, while $\sim55\pm10$\% SWEPs are metal-enriched, and only
$53\pm11$\% stars with huge planets/brown dwarfs are metal-enriched. Again, the
fraction $f$ shows a clear turn-off pattern, which peaks at Jupiter size as
well. 

These turn-off patterns have not been reported yet, as the exercise extending
companion's radius to far beyond Jupiter's radius has not been conducted in the
past. We are aware those very large companions are probably not planets, and
the purpose to include them in the current analysis is mainly to help our
understanding of formation of planets larger than the Jupiter, and the
difference of the giant plant formation and brown dwarf formation. This
overall trend confirms that the P-M correlation is strong and positive for
Jupiter-size planets, and is weakened for Neptune-size planets
\citep{FV2005,Sousa2011}, and however contrary to the conclusions from some
other works \citep[e.g.  ][]{WF2015}. On the other hand, this also suggests
that the P-M correlation might have disappeared for Earth-sized planets and for
much bigger planets/brown dwarfs. In summary, the P-M correlation is dependent
on planet radius $R_{\rm P}$.

We admit that the KSNP sample is not clean, because only those planets with
small inclination angles can be detected with the transit method. Therefore in
the KSNP sample, there should be noticeable ``false'' SNP stars actually
hosting planets. To give a rough estimate of the False Alarm Possibility of
SWPs in the KSNP sample, we adopt the average number of planets per star for
different periods ranges for per planet size bin listed in Table 3 of
~\cite{Fressin2013}. We estimate that for stars with large planets ($R_{\rm
P}>4R_{\oplus}$), the fraction of SWPs in the KSNP sample is no more than 6
percent. Taking into account a maximum metal enrichment of $\sim$0.2\,dex for
large planets, they should not increase the mean metallicities in each radius
bin by more than $0.012$\,dex. For stars with small planets ($R_{\rm
P}<4R_{\oplus}$), this FAP can be as high as 31\%. However, the resulting
increase of mean metallicities should be smaller than 0.016\,dex, given the
currently measured metal enrichment of no more than $\sim0.05$\,dex.
Considering that both 0.012\,dex and 0.016\,dex are smaller than the error bars
given in Fig.~\ref{fig:FeH_vs_Teff}, \ref{fig:dfeh2prad} \&
\ref{fig:dfeh2prad_FGK}, the pattern obtained from the uncleaned KSNP sample
remain in principal untouched.

We point out that for the SWP sample, most planets are identified by the
{\textit Kepler} mission as planet candidates, or referred to as {\textit
Kepler} Objects of Interests (KOIs) from their periodic transit-like light
curves. Previous investigations including, for example, \citet{Fressin2013} and
\cite{Morton2016} show that the FAP (False Alarm Possibility) is about 7-9\% for
Kepler candidates with $R_{\rm P}<4R_\oplus$, and is about 16-22\% for
candidates with $R_{\rm P}>4R_\oplus$. Which means that for KOIs, $\sim$8\% or
19\% of them are actually not planetary systems, and they should be
re-classified as stars without planets. For SWPs with $R_{\rm P}>2R_\oplus$,
the P-M correlation is observed to be positive, and therefore wrongly including
SNPs in our SWP sample will only make the metal enhancement less significant by
$\sim$20\%, or $\sim$ 0.1\,dex. For the case of SWEPs, i.e., $R_{\rm P}<2R_\oplus
$, the extent of over-metallicity is not significant, and therefore we can
not tell which direction the false alarms would affect, but given the FAP of
8\%, the resulting bias should not be noticeable. 

\subsection{The P-M correlation is found to depend on stellar metallicity!}

\begin{figure}
\centering
\includegraphics[width=9.5cm,clip=true, bb=15 8 610 550]{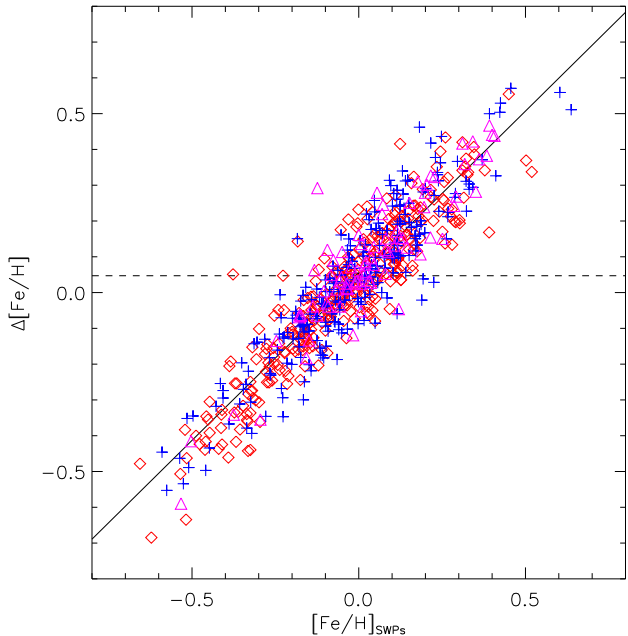}
\caption{\label{fig:dfeh2feh}$\Delta$[Fe/H] against [Fe/H]$_{\rm SWP}$. The
solid line represents the best-fit curve with a slope of $0.93\pm0.02$ and
intercept of $0.05\pm0.004$ respectively. The dashed line shows the intercept
of the best fit. Red diamonds, blue crosses and purple triangles represent
SWEPs, SWNPs and SWJPs, respectively.}
\end{figure}

It has been investigated and discussed for a long time about the dependence of
the P-M correlation on $R_{\rm P}$. However, it has never been reported that
the P-M correlation also depends on other stellar or planetary parameters.
When playing with our SWP-CKSNP comparison sample, we find it quite
interesting from Fig.\ref{fig:dfeh2feh} that $\Delta$[Fe/H] shows a positive
linear correlation with [Fe/H]$_{\rm SWP}$ with a slope of $0.93\pm0.02$ and
intercept of $0.05\pm0.004$, which has never been reported previously. The
non-unity slope and marginally positive intercept support the fact that SWP is
statistically slightly metal rich than CKSNP at solar metallicity, and the
metal enhancement is quite significant at super-solar metallicity and becomes
negative at subsolar metallicities. This suggests that the so-called P-M
relation reported previously is in fact metallicity-dependent, and it is only
valid for solar and super-solar metallicity stars. We realize that this
relationship also holds if the whole sample is divided into three subsamples with
different sizes of planets. Linear regressions of the SWJPs (purple triangles),
SWNP (blue crosses) and SWEP (red diamonds) subsamples yield quite similar
correlation coefficients, ranging from 0.92 to 0.95 for the slopes, and $0.04$
to $0.06$ for the intercepts. We note SWJPs are mostly having higher [Fe/H] and
SWNPs with median values, while SWEPs with lower values.

\begin{figure*}
\includegraphics[width=18cm,clip=true, bb=0 0 612 282]{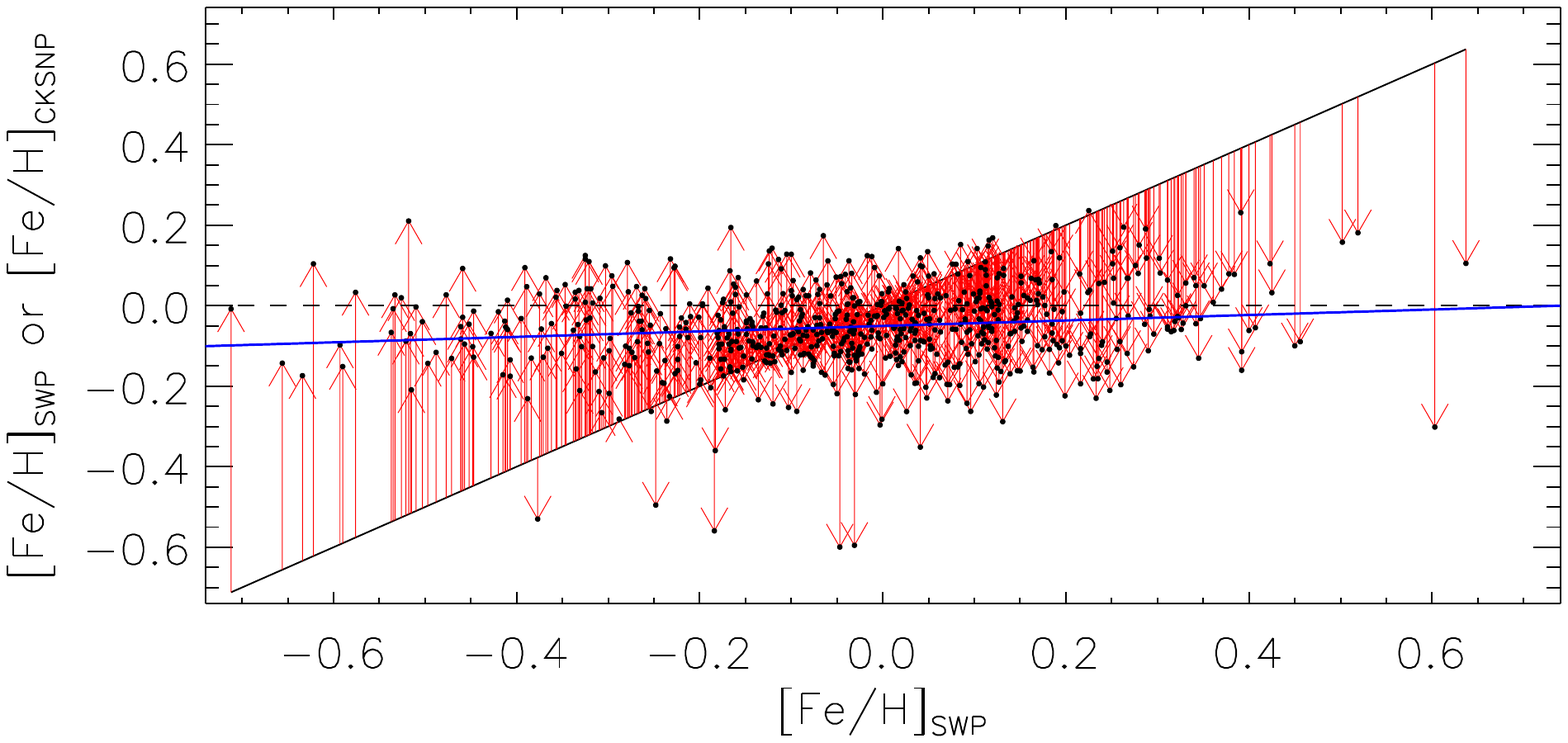}
\caption{\label{fig:feh2feh}[Fe/H]$_{\rm SWP}$ (black line) and [Fe/H]$_{\rm
CKSNP}$ (black dots) against [Fe/H]$_{\rm SWP}$. The arrows show the directions
from [Fe/H]$_{\rm SWP}$ to [Fe/H]$_{\rm CKSNP}$ for each pair. The blue
straight line is the best linear fit to [Fe/H]$_{\rm CKSNP}$ against
[Fe/H]$_{\rm SWP}$. }
\end{figure*}

The linear relationship might be a mathematical result from the fact that
[Fe/H]$_{\rm SWP}$ has a larger dispersion ($\sim$0.21\,dex), as compared to
that of [Fe/H]$_{\rm CKSNP}$ ($\sim$0.10\,dex), as clearly shown in
Fig.~\ref{fig:feh2feh} and Fig.~\ref{fig:FeH_dist}. In the former figure,
[Fe/H]$_{\rm SWP}$ (the black connected line) and [Fe/H]$_{\rm CKSNP}$ (the
black scatter dots) are plotted one by one, as a function of the [Fe/H]$_{\rm
SWP}$ . It is clear that [Fe/H]$_{\rm SWP}$ increases steadily from about
$-0.7$ to 0.65\,dex, while [Fe/H]$_{\rm CKSNP}$ is nearly flat around zero
metallicity, and is mostly between $-0.25$ and 0.20\,dex. The blue line is the
least-square linear fit to [Fe/H]$_{\rm CKSNP}$ as a weak function of
[Fe/H]$_{\rm SWP}$, giving a hint that the former increases slowly with the
latter. To make sure that this conclusion is not affected by data
uncertainties, we create a new set of [Fe/H]$_{\rm CKSNP}$ by increasing the
deviations of [Fe/H]$_{\rm CKSNP}$ around their local average values by a
factor of two. Then we perform similar regression exercise to the new set of
[Fe/H]$_{\rm CKSNP}$, and we find similar although weaker linear correlations,
which suggests that the observed linear relationship is more likely to be real.

No matter whether this linear relationship is a mathematical consequence or
not, the key point and more fundamental character shown in our samples is that
[Fe/H]$_{\rm SWP}$ has a larger variation, while [Fe/H]$_{\rm CKSNP}$ does not.
We point out that as all the CKSNP stars are selected from within the Kepler
field, it is quite reasonable that their metallicities are narrowly
distributed. On the other hand, SWP stars have metallicities higher than their
counterparts by various extent mainly related to various $R_{\rm P}$ and host
stars' spectral type, and therefore have significantly larger variations. This
obvious difference in the variations of [Fe/H] for SWPs and SNPs, and the
strong correlation of $\Delta$[Fe/H] against [Fe/H]$_{\rm SWPs}$ evidently
suggest that the Planet-Metallicity correlation (if exists) is
metallcity-dependent, which is positive at high metallicities (solar and super
solar), and becomes negative for subsolar metallicites. 

We note that $\Delta$[Fe/H] are dependent both on [Fe/H] and on $R_{\rm P}$ is
not a surprise, given the fact that most SWJPs which have large $\Delta$[Fe/H]
locate mostly in the upper-right portion in Fig.\ref{fig:dfeh2feh}. This
turn-off trends of $\Delta$[Fe/H] and $f$ against $R_{\rm P}$ indicates that
the gas giant planets are different from rocky planets and brown dwarfs, in the
aspect of the metallicities of their host stars, and thus implies a different
mechanisms and requirements for the formation of gas giants, compared to rocky
planets, and to more massive planets/brown dwarfs. This implication naturally
excludes the disk instability scenario for gas giant formation. 

We should make another statement that, the metal pollution scenario, i.e., host
stars engulfing migrating planets into their photospheres and therefore enrich
their apparent metallicity, can not be the only reason for the P-M correlation
for SWJPs. Metallic material of $5-10M{\oplus}$ from rocky planets or giant
planets are believed to be enough to enrich the apparent photospheric
metallicities by about 0.1\,dex for G-type stars, more for earlier type stars
and less for later type stars, depending on the mass of their photospheric
layers \citep{Gonzalez1997}. In this scenario, massive rocky planets should have
similar although less-serious pollution, and $\Delta$[Fe/H] should increases
with $R_{\rm P}$ because larger planets should contain more metals. The first
four red and purple dots in Fig.~\ref{fig:dfeh2prad} are almost identical
within error bars, and there is only marginal ascending trend observable and
therefore our data set does not support this metal pollution scenario. More
accurate measurements of [Fe/H] should help to better constrain this trend for
our sample. 

Moreover, we argue that the higher metallicity is a requirement for gas giants
to form, but not necessary for rocky planets and brown dwarfs. This is
supported by the obvious deficient of SWJPs in the blue box region in
Fig.~\ref{fig:dfeh2prad}, and the high $f$ of SWJPs and SWNPs. Just to mention,
the 7 SWPs that have modestly low metallicity worth future high-resolution
study, which might be helpful for the understanding of how planets form in
metal-poor environments. We note that it is not necessary at all to be
metal-rich to form small planets, arguing against the theoretical requirement
of [Fe/H]$>0.3$ to bear terrestrial planets \citep{Gonzalez2001}. In other
words, the birthplace of terrestrial planets might be widespread in the Galaxy,
and even possible back to the early Universe, as pointed out for example by
\citet{Buchhave2012}, and \citet{Zackrisson2016}.

\subsection{The P-M correlation is dependent on the stellar spectral type}

\begin{figure}
\centering
\includegraphics[width=9.5cm,clip=true, bb=15 5 612 600]{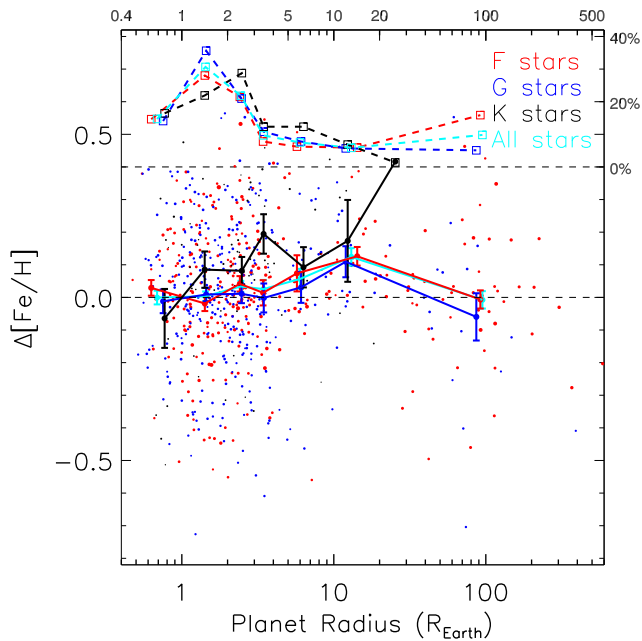}

\caption{\label{fig:dfeh2prad_FGK} Same data and same labels as
Fig.\ref{fig:dfeh2prad}, but with stars of F, G, K and all spectral type
differentially colored in red, blue, black and cyan, respectively. The
connected filled circles with error bars are the mean values and standard
errors in the same $R_{\rm P}$ bins. The most right black dot does not have
error bar because there is only one object in this bin for K stars. For
each spectral type from F to K, the fractions of stars in each $R_{\rm P}$ bin
with respect to the total number of stars in each spectral type are shown as
open squares connected by dashed lines with the same color coding as in the
upper part of the plot. The upper horizontal dashed line represents where the
fractions equal zero percent, while the lower one represents where
$\Delta$[Fe/H] equals zero.} 

\end{figure}

We show in Fig.~\ref{fig:dfeh2prad_FGK} the same data but with stars of F, G
and K spectral types color-coded in red, blue and black symbols, respectively,
both in the scatter plot and in the average trend (the filled circles with
error bars). It is clear that F and G-type stars behave quite similar, while
K-type stars are different, in the sense that they have higher 
($\sim2\sigma$) $\Delta$[Fe/H] for SWNPs (the forth bin in
Fig.~\ref{fig:dfeh2prad_FGK}), and are marginally (about or slightly less than
$1\sigma$) higher for SWEPs (the second and third bins). If the observed
difference is real, we argue that it might originate from the different
convective zones in the scenario of pollution, or is due to different mass of
proto-planetary disk in the scenario of core-accretion scenario. This is
consistent with the implication presented by the approach of the solid and
dashed line in the high mass end in the middle panel of Fig.~2 in
\citet{Johnson2010}, which hints for a decreasing $\Delta$[Fe/H] with stellar
mass increasing.

Stellar models have been constructed by \cite{Pinsonneault2001} to estimate the
mass of the outer convective zone in FGK main sequence (MS) stars, and the mass
of convective zone is found to decreases dramatically with stellar mass, and
any contamination of a star's atmosphere by accreted planetary material should
affect hotter stars much more than cool stars. This trend is not observed in
this plot, however. In fact, the trend shown in this plot is contrary to this
prediction and therefore it strongly rejects the pollution scenario for the P-M
correlation. In addition, the significant high metallicity of K-type SWPs with
$R_{\rm P}>3R_{\oplus}$ suggest that K stars requires much higher content of
metals to form large planets in their discs than FG stars, which can be
naturally explained by the core-accretion model as less massive disk needs more
metals to form a rock core more massive than approximately 10\,$M_{\oplus}$. 

In addition, we show in the upper portion of Fig.~\ref{fig:dfeh2prad_FGK}
in units of percentage the ratios of the number of stars in each radius bin of
a certain spectral type to the total number of stars of the same spectral type.
It is clear and quite interesting as shown that F and G-type stars again have
similar distributions, while K-type stars are significantly less frequent for
Earth-sized planets, and turns to be more frequent at Neptune-sized planets.
We will discuss the difference in the occurrence rate for SWPs of different spectral
types in our next work (Wang et al. {\it in prep.}), and in this paper the
distributions are only used to exclude the possibility that the differences in
metal enhancement pattern for FG-type stars and K-type stars result from the
differences in frequencies, because they (the solid and broken lines) have
quite different trends. We note that the numbers of SWPs of F, G and
K-spectral types are 329, 336 and 73, respectively. This is the main reason
that the overall trend for the entire sample (the cyan lines) follow those of F
and G SWPs (the red and blue curves).

\section{conclusion}

We employ the LMDR3 stellar spectral library with the Kepler planet candidates
to create the largest samples of stars with planets in a relatively restricted
volume in the Milky Way, and a well-defined sample of stars with no planets,
with which we perform detailed investigations of the P-M correlation. We
firstly point out that the selection of the control sample (the SNP sample) is
very crucial, given that most of them are in the Milky way which shows
metallicity gradient, and therefore stars at different positions inherently
contain different metallicities. We show that with control samples averagely
further away from us comparing to the target sample (the SWP sample), they have
less metallicities gradually. For our study, by constraining the $T_{\rm eff}$,
log\,$g$, $r$ magnitude and $g-r$ color of the SNPs to be close to those of the
SWPs, the comparison pair should have similar stellar properties (excluding
metallcities) and distances from the Sun. We believe our selected control
sample is the best for such kind of statistical studies. 

We find that in general SWPs are more metal-rich than SNPs, but due to the fact
that most of our SWPs' planets are Earth- Size or Neptune-size, and the extent
of the over-metallicity of our SWP sample as compared to the SNP sample is not
significant. We confirm that the P-M correlation, or the higher-metallicity of
SWPs compared to SNPs, is strong for SWJPs, and weaker for SWNPs, and become
only marginal for SWEPs and SWMPs (star with massive planets). In other words,
we find for the first time a clear peak of $\Delta$[Fe/H] at Jupiter-sized
planets. In addition, the percentage of stars with $\Delta$[Fe/H]$>0$ has a
similar trend. Based on this turn-over trend, we believe that the mechanism of
giant planets formation might be different from both rocky planets and very
massive planets or brown dwarfs, so they should not form through only
core-collapse or solely through collisional growth of planet embryos. 

We also find it obvious that FG type stars follow this general trend, but K
stars are slightly different in the sense that obvious metal enhancement
($\sim$0.1-0.2\,dex) is detected for stars bearing Neptuian-sized planets
and Super-Earths as compared to SNPs, which indicates that much higher
metallcities are required for Neptunes and Super-Earths to form around K type
stars, than those around FG stars.

We point out for the first time that the P-M correlation is also dependent on
stellar metallicities, i.e., we observe positive $\Delta$[Fe/H] at solar or
super-solar metallicities, and negative $\Delta$[Fe/H] at subsolar
metallicities. The dependence may be mathematically explained by the fact the
SWPs has a metallicity deviation about twice of SNPs', and there when
[Fe/H]$_{\rm SWPs}$ increases or decreases, $\Delta$[Fe/H] increases and
decreases accordingly, but with a slightly less extent. 

To summary, we conclude that giant planets probably form differently from rocky
planets or more massive planets/brown dwarfs, and the core-accretion scenario
is highly favored by our result. No steady increase of $\Delta$[Fe/H] within
error bars against planet sizes is observed for stars with rocky planets, which
could rule out the metal pollution theory for the apparent P-M correlation,
because massive rocky planets should also be able to contribute a large amount
of metals into stellar photospheres if engulfed. Instead, this observation
implies a higher metallicity is prerequisite for massive planets to form. 

\acknowledgments

This research is supported by the National Natural Science Foundation of China
(NSFC) under grants No. 11203035 and 11390371, and is accomplished (in part) by
the Chinese Academy of Sciences (CAS), through a grant to the CAS South America
Center for Astronomy (CASSACA) in Santiago, Chile. Y.Q.C. acknowledges support
from NSFC grant No.11625313. 

The Guoshoujing Telescope (the Large Sky Area Multi-Object Fiber Spectroscopic
Telescope, LAMOST) is a National Major Scientific Project built by the Chinese
Academy of Sciences. Funding for the project has been provided by the National
Development and Reform Commission. LAMOST is operated and managed by the
National Astronomical Observatories, Chinese Academy of Sciences. 

\bibliography{planet_atmosphere}

\begin{thebibliography}{56}
\expandafter\ifx\csname natexlab\endcsname\relax\def\natexlab#1{#1}\fi

\bibitem[{{Adibekyan} {et~al.}(2012){Adibekyan}, {Sousa}, {Santos}, {Delgado
  Mena}, {Gonz{\'a}lez Hern{\'a}ndez}, {Israelian}, {Mayor}, \&
  {Khachatryan}}]{Adibekyan2012a}
{Adibekyan}, V.~Z., {Sousa}, S.~G., {Santos}, N.~C., {Delgado Mena}, E.,
  {Gonz{\'a}lez Hern{\'a}ndez}, J.~I., {Israelian}, G., {Mayor}, M., \&
  {Khachatryan}, G. 2012, \aap, 545, A32

\bibitem[{{Batalha} {et~al.}(2013){Batalha}, {Rowe}, {Bryson}, {Barclay},
  {Burke}, {Caldwell}, {Christiansen}, {Mullally}, {Thompson}, {Brown},
  {Dupree}, {Fabrycky}, {Ford}, {Fortney}, {Gilliland}, {Isaacson}, {Latham},
  {Marcy}, {Quinn}, {Ragozzine}, {Shporer}, {Borucki}, {Ciardi}, {Gautier},
  {Haas}, {Jenkins}, {Koch}, {Lissauer}, {Rapin}, {Basri}, {Boss}, {Buchhave},
  {Carter}, {Charbonneau}, {Christensen-Dalsgaard}, {Clarke}, {Cochran},
  {Demory}, {Desert}, {Devore}, {Doyle}, {Esquerdo}, {Everett}, {Fressin},
  {Geary}, {Girouard}, {Gould}, {Hall}, {Holman}, {Howard}, {Howell},
  {Ibrahim}, {Kinemuchi}, {Kjeldsen}, {Klaus}, {Li}, {Lucas}, {Meibom},
  {Morris}, {Pr{\v s}a}, {Quintana}, {Sanderfer}, {Sasselov}, {Seader},
  {Smith}, {Steffen}, {Still}, {Stumpe}, {Tarter}, {Tenenbaum}, {Torres},
  {Twicken}, {Uddin}, {Van Cleve}, {Walkowicz}, \& {Welsh}}]{Batalha2013}
{Batalha}, N.~M., {et~al.} 2013, \apjs, 204, 24

\bibitem[{{Borucki} {et~al.}(2011){Borucki}, {Koch}, {Basri}, {Batalha},
  {Brown}, {Bryson}, {Caldwell}, {Christensen-Dalsgaard}, {Cochran}, {DeVore},
  {Dunham}, {Gautier}, {Geary}, {Gilliland}, {Gould}, {Howell}, {Jenkins},
  {Latham}, {Lissauer}, {Marcy}, {Rowe}, {Sasselov}, {Boss}, {Charbonneau},
  {Ciardi}, {Doyle}, {Dupree}, {Ford}, {Fortney}, {Holman}, {Seager},
  {Steffen}, {Tarter}, {Welsh}, {Allen}, {Buchhave}, {Christiansen}, {Clarke},
  {Das}, {D{\'e}sert}, {Endl}, {Fabrycky}, {Fressin}, {Haas}, {Horch},
  {Howard}, {Isaacson}, {Kjeldsen}, {Kolodziejczak}, {Kulesa}, {Li}, {Lucas},
  {Machalek}, {McCarthy}, {MacQueen}, {Meibom}, {Miquel}, {Prsa}, {Quinn},
  {Quintana}, {Ragozzine}, {Sherry}, {Shporer}, {Tenenbaum}, {Torres},
  {Twicken}, {Van Cleve}, {Walkowicz}, {Witteborn}, \& {Still}}]{Borucki2011}
{Borucki}, W.~J., {et~al.} 2011, \apj, 736, 19

\bibitem[{{Bouchy} {et~al.}(2009){Bouchy}, {Mayor}, {Lovis}, {Udry}, {Benz},
  {Bertaux}, {Delfosse}, {Mordasini}, {Pepe}, {Queloz}, \&
  {Segransan}}]{Bouchy2009}
{Bouchy}, F., {et~al.} 2009, \aap, 496, 527

\bibitem[{{Brown} {et~al.}(2011){Brown}, {Latham}, {Everett}, \&
  {Esquerdo}}]{Brown2011}
{Brown}, T.~M., {Latham}, D.~W., {Everett}, M.~E., \& {Esquerdo}, G.~A. 2011,
  \aj, 142, 112

\bibitem[{{Bruntt} {et~al.}(2012){Bruntt}, {Basu}, {Smalley}, {Chaplin},
  {Verner}, {Bedding}, {Catala}, {Gazzano}, {Molenda-{\.Z}akowicz}, {Thygesen},
  {Uytterhoeven}, {Hekker}, {Huber}, {Karoff}, {Mathur}, {Mosser},
  {Appourchaux}, {Campante}, {Elsworth}, {Garc{\'{\i}}a}, {Handberg},
  {Metcalfe}, {Quirion}, {R{\'e}gulo}, {Roxburgh}, {Stello},
  {Christensen-Dalsgaard}, {Kawaler}, {Kjeldsen}, {Morris}, {Quintana}, \&
  {Sanderfer}}]{Bruntt2012}
{Bruntt}, H., {et~al.} 2012, \mnras, 423, 122

\bibitem[{{Buchhave} \& {Latham}(2015)}]{BL2015}
{Buchhave}, L.~A., \& {Latham}, D.~W. 2015, \apj, 808, 187

\bibitem[{{Buchhave} {et~al.}(2012){Buchhave}, {Latham}, {Johansen},
  {Bizzarro}, {Torres}, {Rowe}, {Batalha}, {Borucki}, {Brugamyer}, {Caldwell},
  {Bryson}, {Ciardi}, {Cochran}, {Endl}, {Esquerdo}, {Ford}, {Geary},
  {Gilliland}, {Hansen}, {Isaacson}, {Laird}, {Lucas}, {Marcy}, {Morse},
  {Robertson}, {Shporer}, {Stefanik}, {Still}, \& {Quinn}}]{Buchhave2012}
{Buchhave}, L.~A., {et~al.} 2012, \nat, 486, 375

\bibitem[{{Buchhave} {et~al.}(2014){Buchhave}, {Bizzarro}, {Latham},
  {Sasselov}, {Cochran}, {Endl}, {Isaacson}, {Juncher}, \&
  {Marcy}}]{Buchhave2014}
---. 2014, \nat, 509, 593

\bibitem[{{Burke} {et~al.}(2013){Burke}, {Bryson}, {Christiansen}, {Mullally},
  {Rowe}, {Science Office}, \& {Kepler Science Team}}]{Burke2013}
{Burke}, C.~J., {Bryson}, S., {Christiansen}, J., {Mullally}, F., {Rowe}, J.,
  {Science Office}, K., \& {Kepler Science Team}. 2013, in American
  Astronomical Society Meeting Abstracts, Vol. 221, American Astronomical
  Society Meeting Abstracts, 216.02

\bibitem[{{Catanzarite} \& {Shao}(2011)}]{CS2011}
{Catanzarite}, J., \& {Shao}, M. 2011, \apj, 738, 151

\bibitem[{{Chaplin} {et~al.}(2014){Chaplin}, {Basu}, {Huber}, {Serenelli},
  {Casagrande}, {Silva Aguirre}, {Ball}, {Creevey}, {Gizon}, {Handberg},
  {Karoff}, {Lutz}, {Marques}, {Miglio}, {Stello}, {Suran}, {Pricopi},
  {Metcalfe}, {Monteiro}, {Molenda-{\.Z}akowicz}, {Appourchaux},
  {Christensen-Dalsgaard}, {Elsworth}, {Garc{\'{\i}}a}, {Houdek}, {Kjeldsen},
  {Bonanno}, {Campante}, {Corsaro}, {Gaulme}, {Hekker}, {Mathur}, {Mosser},
  {R{\'e}gulo}, \& {Salabert}}]{Chaplin2014}
{Chaplin}, W.~J., {et~al.} 2014, \apjs, 210, 1

\bibitem[{{Coelho} {et~al.}(2005){Coelho}, {Barbuy}, {Mel{\'e}ndez},
  {Schiavon}, \& {Castilho}}]{Coelho2005}
{Coelho}, P., {Barbuy}, B., {Mel{\'e}ndez}, J., {Schiavon}, R.~P., \&
  {Castilho}, B.~V. 2005, \aap, 443, 735

\bibitem[{{Cui} {et~al.}(2012){Cui}, {Zhao}, {Chu}, {Li}, {Li}, {Zhang}, {Su},
  {Yao}, {Wang}, {Xing}, {Li}, {Zhu}, {Wang}, {Gu}, {Luo}, {Xu}, {Zhang},
  {Liu}, {Zhang}, {Yang}, {Cao}, {Chen}, {Chen}, {Chen}, {Chen}, {Chu}, {Feng},
  {Gong}, {Hou}, {Hu}, {Hu}, {Hu}, {Jia}, {Jiang}, {Jiang}, {Jiang}, {Jin},
  {Li}, {Li}, {Li}, {Liu}, {Liu}, {Lu}, {Mao}, {Men}, {Qi}, {Qi}, {Shi},
  {Tang}, {Tao}, {Wang}, {Wang}, {Wang}, {Wang}, {Wang}, {Wang}, {Wang},
  {Wang}, {Wang}, {Wang}, {Wang}, {Wang}, {Xu}, {Xu}, {Yang}, {Yu}, {Yuan},
  {Yuan}, {Zhai}, {Zhang}, {Zhang}, {Zhang}, {Zhao}, {Zhou}, {Zhou}, {Zhu}, \&
  {Zou}}]{Cui2012}
{Cui}, X.-Q., {et~al.} 2012, Research in Astronomy and Astrophysics, 12, 1197

\bibitem[{{De Cat} {et~al.}(2015){De Cat}, {Fu}, {Ren}, {Yang}, {Shi}, {Luo},
  {Yang}, {Wang}, {Zhang}, {Shi}, {Zhang}, {Dong}, {Catanzaro}, {Corbally},
  {Frasca}, {Gray}, {Molenda-{\.Z}akowicz}, {Uytterhoeven}, {Briquet},
  {Bruntt}, {Frandsen}, {Kiss}, {Kurtz}, {Marconi}, {Niemczura}, {{\O}stensen},
  {Ripepi}, {Smalley}, {Southworth}, {Szab{\'o}}, {Telting}, {Karoff}, {Silva
  Aguirre}, {Wu}, {Hou}, {Jin}, \& {Zhou}}]{DeCat2015}
{De Cat}, P., {et~al.} 2015, \apjs, 220, 19

\bibitem[{{Deng} {et~al.}(2012){Deng}, {Newberg}, {Liu}, {Carlin}, {Beers},
  {Chen}, {Chen}, {Christlieb}, {Grillmair}, {Guhathakurta}, {Han}, {Hou},
  {Lee}, {L{\'e}pine}, {Li}, {Liu}, {Pan}, {Sellwood}, {Wang}, {Wang}, {Yang},
  {Yanny}, {Zhang}, {Zhang}, {Zheng}, \& {Zhu}}]{Deng2012}
{Deng}, L.-C., {et~al.} 2012, Research in Astronomy and Astrophysics, 12, 735

\bibitem[{{Dong} {et~al.}(2018){Dong}, {Xie}, {Zhou}, {Zheng}, \&
  {Luo}}]{Dong2018}
{Dong}, S., {Xie}, J.-W., {Zhou}, J.-L., {Zheng}, Z., \& {Luo}, A. 2018,
  Proceedings of the National Academy of Science, 115, 266

\bibitem[{{Dong} {et~al.}(2014){Dong}, {Zheng}, {Zhu}, {De Cat}, {Fu}, {Yang},
  {Zhang}, {Jin}, \& {Zhang}}]{Dong2014}
{Dong}, S., {et~al.} 2014, \apjl, 789, L3

\bibitem[{{Dressing} \& {Charbonneau}(2013)}]{DC2013}
{Dressing}, C.~D., \& {Charbonneau}, D. 2013, \apj, 767, 95

\bibitem[{{Ercolano} \& {Clarke}(2010)}]{EC2010}
{Ercolano}, B., \& {Clarke}, C.~J. 2010, \mnras, 402, 2735

\bibitem[{{Everett} {et~al.}(2013){Everett}, {Howell}, {Silva}, \&
  {Szkody}}]{Everett2013}
{Everett}, M.~E., {Howell}, S.~B., {Silva}, D.~R., \& {Szkody}, P. 2013, \apj,
  771, 107

\bibitem[{{Fischer} \& {Valenti}(2005)}]{FV2005}
{Fischer}, D.~A., \& {Valenti}, J. 2005, \apj, 622, 1102

\bibitem[{{Fressin} {et~al.}(2013){Fressin}, {Torres}, {Charbonneau}, {Bryson},
  {Christiansen}, {Dressing}, {Jenkins}, {Walkowicz}, \&
  {Batalha}}]{Fressin2013}
{Fressin}, F., {et~al.} 2013, \apj, 766, 81

\bibitem[{{Ghezzi} {et~al.}(2010){Ghezzi}, {Cunha}, {Schuler}, \&
  {Smith}}]{Ghezzi2010}
{Ghezzi}, L., {Cunha}, K., {Schuler}, S.~C., \& {Smith}, V.~V. 2010, \apj, 725,
  721

\bibitem[{{Gonzalez}(1997)}]{Gonzalez1997}
{Gonzalez}, G. 1997, \mnras, 285, 403

\bibitem[{{Gonzalez} {et~al.}(2001){Gonzalez}, {Brownlee}, \&
  {Ward}}]{Gonzalez2001}
{Gonzalez}, G., {Brownlee}, D., \& {Ward}, P. 2001, Icarus, 152, 185

\bibitem[{{Han} {et~al.}(2014){Han}, {Wang}, {Wright}, {Feng}, {Zhao},
  {Fakhouri}, {Brown}, \& {Hancock}}]{Han2014}
{Han}, E., {Wang}, S.~X., {Wright}, J.~T., {Feng}, Y.~K., {Zhao}, M.,
  {Fakhouri}, O., {Brown}, J.~I., \& {Hancock}, C. 2014, \pasp, 126, 827

\bibitem[{{Howard} {et~al.}(2012){Howard}, {Marcy}, {Bryson}, {Jenkins},
  {Rowe}, {Batalha}, {Borucki}, {Koch}, {Dunham}, {Gautier}, {Van Cleve},
  {Cochran}, {Latham}, {Lissauer}, {Torres}, {Brown}, {Gilliland}, {Buchhave},
  {Caldwell}, {Christensen-Dalsgaard}, {Ciardi}, {Fressin}, {Haas}, {Howell},
  {Kjeldsen}, {Seager}, {Rogers}, {Sasselov}, {Steffen}, {Basri},
  {Charbonneau}, {Christiansen}, {Clarke}, {Dupree}, {Fabrycky}, {Fischer},
  {Ford}, {Fortney}, {Tarter}, {Girouard}, {Holman}, {Johnson}, {Klaus},
  {Machalek}, {Moorhead}, {Morehead}, {Ragozzine}, {Tenenbaum}, {Twicken},
  {Quinn}, {Isaacson}, {Shporer}, {Lucas}, {Walkowicz}, {Welsh}, {Boss},
  {Devore}, {Gould}, {Smith}, {Morris}, {Prsa}, {Morton}, {Still}, {Thompson},
  {Mullally}, {Endl}, \& {MacQueen}}]{Howard2012}
{Howard}, A.~W., {et~al.} 2012, \apjs, 201, 15

\bibitem[{{Ida} \& {Lin}(2004)}]{IL2004}
{Ida}, S., \& {Lin}, D.~N.~C. 2004, \apj, 616, 567

\bibitem[{{Ida} \& {Lin}(2005)}]{IL2005}
---. 2005, \apj, 626, 1045

\bibitem[{{Johnson} {et~al.}(2010){Johnson}, {Aller}, {Howard}, \&
  {Crepp}}]{Johnson2010}
{Johnson}, J.~A., {Aller}, K.~M., {Howard}, A.~W., \& {Crepp}, J.~R. 2010,
  \pasp, 122, 905

\bibitem[{{Lin} {et~al.}(1996){Lin}, {Bodenheimer}, \& {Richardson}}]{Lin1996}
{Lin}, D.~N.~C., {Bodenheimer}, P., \& {Richardson}, D.~C. 1996, Nature, 380,
  606

\bibitem[{{Lissauer} {et~al.}(2012){Lissauer}, {Marcy}, {Rowe}, {Bryson},
  {Adams}, {Buchhave}, {Ciardi}, {Cochran}, {Fabrycky}, {Ford}, {Fressin},
  {Geary}, {Gilliland}, {Holman}, {Howell}, {Jenkins}, {Kinemuchi}, {Koch},
  {Morehead}, {Ragozzine}, {Seader}, {Tanenbaum}, {Torres}, \&
  {Twicken}}]{Lissauer2012}
{Lissauer}, J.~J., {et~al.} 2012, \apj, 750, 112

\bibitem[{{Luo} {et~al.}(2012){Luo}, {Zhang}, {Zhao}, {Zhao}, {Cui}, {Li},
  {Chu}, {Shi}, {Wang}, {Zhang}, {Bai}, {Chen}, {Wang}, {Guo}, {Chen}, {Du},
  {Kong}, {Lei}, {Li}, {Song}, {Wu}, {Zhang}, {Zhou}, {Zuo}, {Du}, {He}, {Hou},
  {Dong}, {Li}, {Li}, {Li}, {Song}, {Tian}, {Wang}, {Wu}, {Yang}, {Yuan},
  {Cao}, {Chen}, {Chen}, {Chen}, {Chu}, {Feng}, {Gong}, {Gu}, {Hou}, {Huo},
  {Hu}, {Hu}, {Hu}, {Jia}, {Jiang}, {Jiang}, {Jiang}, {Jin}, {Li}, {Li}, {Li},
  {Li}, {Li}, {Liu}, {Liu}, {Liu}, {Lu}, {Lu}, {Luo}, {Mao}, {Men}, {Ni}, {Qi},
  {Qi}, {Shi}, {Su}, {Sun}, {Su}, {Tang}, {Tao}, {Tu}, {Wang}, {Wang}, {Wang},
  {Wang}, {Wang}, {Wang}, {Wang}, {Wang}, {Wang}, {Wang}, {Wang}, {Wang},
  {Wang}, {Wang}, {Wei}, {Xue}, {Xing}, {Xu}, {Xu}, {Xu}, {Yang}, {Yang},
  {Yao}, {Yu}, {Yuan}, {Zhai}, {Zhang}, {Zhang}, {Zhang}, {Zhang}, {Zhang},
  {Zhang}, {Zhao}, {Zhou}, {Zhu}, {Zhu}, \& {Zou}}]{Luo2012}
{Luo}, A.-L., {et~al.} 2012, Research in Astronomy and Astrophysics, 12, 1243

\bibitem[{{Marcy} {et~al.}(2014){Marcy}, {Isaacson}, {Howard}, {Rowe},
  {Jenkins}, {Bryson}, {Latham}, {Howell}, {Gautier}, {Batalha}, {Rogers},
  {Ciardi}, {Fischer}, {Gilliland}, {Kjeldsen}, {Christensen-Dalsgaard},
  {Huber}, {Chaplin}, {Basu}, {Buchhave}, {Quinn}, {Borucki}, {Koch}, {Hunter},
  {Caldwell}, {Van Cleve}, {Kolbl}, {Weiss}, {Petigura}, {Seager}, {Morton},
  {Johnson}, {Ballard}, {Burke}, {Cochran}, {Endl}, {MacQueen}, {Everett},
  {Lissauer}, {Ford}, {Torres}, {Fressin}, {Brown}, {Steffen}, {Charbonneau},
  {Basri}, {Sasselov}, {Winn}, {Sanchis-Ojeda}, {Christiansen}, {Adams},
  {Henze}, {Dupree}, {Fabrycky}, {Fortney}, {Tarter}, {Holman}, {Tenenbaum},
  {Shporer}, {Lucas}, {Welsh}, {Orosz}, {Bedding}, {Campante}, {Davies},
  {Elsworth}, {Handberg}, {Hekker}, {Karoff}, {Kawaler}, {Lund}, {Lundkvist},
  {Metcalfe}, {Miglio}, {Silva Aguirre}, {Stello}, {White}, {Boss}, {Devore},
  {Gould}, {Prsa}, {Agol}, {Barclay}, {Coughlin}, {Brugamyer}, {Mullally},
  {Quintana}, {Still}, {Thompson}, {Morrison}, {Twicken}, {D{\'e}sert},
  {Carter}, {Crepp}, {H{\'e}brard}, {Santerne}, {Moutou}, {Sobeck}, {Hudgins},
  {Haas}, {Robertson}, {Lillo-Box}, \& {Barrado}}]{Marcy2014}
{Marcy}, G.~W., {et~al.} 2014, \apjs, 210, 20

\bibitem[{{Molenda-{\.Z}akowicz} {et~al.}(2013){Molenda-{\.Z}akowicz}, {Sousa},
  {Frasca}, {Uytterhoeven}, {Briquet}, {Van Winckel}, {Drobek}, {Niemczura},
  {Lampens}, {Lykke}, {Bloemen}, {Gameiro}, {Jean}, {Volpi}, {Gorlova},
  {Mortier}, {Tsantaki}, \& {Raskin}}]{Molenda2013}
{Molenda-{\.Z}akowicz}, J., {et~al.} 2013, \mnras, 434, 1422

\bibitem[{{Morton} {et~al.}(2016){Morton}, {Bryson}, {Coughlin}, {Rowe},
  {Ravichandran}, {Petigura}, {Haas}, \& {Batalha}}]{Morton2016}
{Morton}, T.~D., {Bryson}, S.~T., {Coughlin}, J.~L., {Rowe}, J.~F.,
  {Ravichandran}, G., {Petigura}, E.~A., {Haas}, M.~R., \& {Batalha}, N.~M.
  2016, \apj, 822, 86

\bibitem[{{Morton} \& {Johnson}(2011)}]{MJ2011}
{Morton}, T.~D., \& {Johnson}, J.~A. 2011, \apj, 738, 170

\bibitem[{{Mulders} {et~al.}(2016){Mulders}, {Pascucci}, {Apai}, {Frasca}, \&
  {Molenda-{\.Z}akowicz}}]{Mulders2016}
{Mulders}, G.~D., {Pascucci}, I., {Apai}, D., {Frasca}, A., \&
  {Molenda-{\.Z}akowicz}, J. 2016, \aj, 152, 187

\bibitem[{{Murray} \& {Chaboyer}(2002)}]{Murray2002b}
{Murray}, N., \& {Chaboyer}, B. 2002, \apj, 566, 442

\bibitem[{{Murray} {et~al.}(2002){Murray}, {Paskowitz}, \&
  {Holman}}]{Murray2002a}
{Murray}, N., {Paskowitz}, M., \& {Holman}, M. 2002, \apj, 565, 608

\bibitem[{{Neves} {et~al.}(2013){Neves}, {Bonfils}, {Santos}, {Delfosse},
  {Forveille}, {Allard}, \& {Udry}}]{Neves2013}
{Neves}, V., {Bonfils}, X., {Santos}, N.~C., {Delfosse}, X., {Forveille}, T.,
  {Allard}, F., \& {Udry}, S. 2013, \aap, 551, A36

\bibitem[{{Pinsonneault} {et~al.}(2001){Pinsonneault}, {DePoy}, \&
  {Coffee}}]{Pinsonneault2001}
{Pinsonneault}, M.~H., {DePoy}, D.~L., \& {Coffee}, M. 2001, \apjl, 556, L59

\bibitem[{{Santerne} {et~al.}(2012){Santerne}, {D{\'{\i}}az}, {Moutou},
  {Bouchy}, {H{\'e}brard}, {Almenara}, {Bonomo}, {Deleuil}, \&
  {Santos}}]{Santerne2012}
{Santerne}, A., {et~al.} 2012, \aap, 545, A76

\bibitem[{{Schlaufman}(2015)}]{Schlaufman2015}
{Schlaufman}, K.~C. 2015, \apjl, 799, L26

\bibitem[{{Soderblom} \& {King}(1998)}]{SK1998}
{Soderblom}, D.~R., \& {King}, J.~R. 1998, in Solar Analogs: Characteristics
  and Optimum Candidates., ed. J.~C. {Hall}, 41

\bibitem[{{Sousa} {et~al.}(2011){Sousa}, {Santos}, {Israelian}, {Mayor}, \&
  {Udry}}]{Sousa2011}
{Sousa}, S.~G., {Santos}, N.~C., {Israelian}, G., {Mayor}, M., \& {Udry}, S.
  2011, \aap, 533, A141

\bibitem[{{Sousa} {et~al.}(2008){Sousa}, {Santos}, {Mayor}, {Udry},
  {Casagrande}, {Israelian}, {Pepe}, {Queloz}, \& {Monteiro}}]{Sousa2008}
{Sousa}, S.~G., {et~al.} 2008, \aap, 487, 373

\bibitem[{{Thygesen} {et~al.}(2012){Thygesen}, {Frandsen}, {Bruntt},
  {Kallinger}, {Andersen}, {Elsworth}, {Hekker}, {Karoff}, {Stello},
  {Brogaard}, {Burke}, {Caldwell}, \& {Christiansen}}]{Thygesen2012}
{Thygesen}, A.~O., {et~al.} 2012, \aap, 543, A160

\bibitem[{{Wang} \& {Fischer}(2013)}]{WF2013}
{Wang}, J., \& {Fischer}, D.~A. 2013, ArXiv e-prints

\bibitem[{{Wang} \& {Fischer}(2015)}]{WF2015}
---. 2015, \aj, 149, 14

\bibitem[{{Wang} {et~al.}(2016){Wang}, {Wang}, {Wu}, {Zhao}, {Li}, {Luo},
  {Liu}, {Zhang}, {Hou}, {Wang}, \& {Cao}}]{WangL2016}
{Wang}, L., {et~al.} 2016, \aj, 152, 6

\bibitem[{{Wu} {et~al.}(2014){Wu}, {Du}, {Luo}, {Zhao}, \& {Yuan}}]{Wu2014}
{Wu}, Y., {Du}, B., {Luo}, A., {Zhao}, Y., \& {Yuan}, H. 2014, in IAU
  Symposium, Vol. 306, Statistical Challenges in 21st Century Cosmology, ed.
  A.~{Heavens}, J.-L. {Starck}, \& A.~{Krone-Martins}, 340--342

\bibitem[{{Wu} {et~al.}(2011){Wu}, {Luo}, {Li}, {Shi}, {Prugniel}, {Liang},
  {Zhao}, {Zhang}, {Bai}, {Wei}, {Dong}, {Zhang}, \& {Chen}}]{Wu2011}
{Wu}, Y., {et~al.} 2011, Research in Astronomy and Astrophysics, 11, 924

\bibitem[{{Zackrisson} {et~al.}(2016){Zackrisson}, {Calissendorff},
  {Gonz{\'a}lez}, {Benson}, {Johansen}, \& {Janson}}]{Zackrisson2016}
{Zackrisson}, E., {Calissendorff}, P., {Gonz{\'a}lez}, J., {Benson}, A.,
  {Johansen}, A., \& {Janson}, M. 2016, \apj, 833, 214

\bibitem[{{Zhao} {et~al.}(2012){Zhao}, {Zhao}, {Chu}, {Jing}, \&
  {Deng}}]{Zhaogang2012}
{Zhao}, G., {Zhao}, Y.-H., {Chu}, Y.-Q., {Jing}, Y.-P., \& {Deng}, L.-C. 2012,
  Research in Astronomy and Astrophysics, 12, 723

\end{thebibliography}

\end{document}